\documentclass[prd,draft,preprint,showpacs,groupedaddress]{revtex4-1}

\usepackage{amsmath}
\usepackage{amsfonts}
\usepackage{amssymb}
\usepackage{geometry}
\usepackage[english]{babel}

\begin{document}
\bibliographystyle{apsrev4-1}
\message{}

  \title{A cosmological model from emergence of space}

  \author{Zi-Liang Wang} \author{Wen-Yuan Ai} \author{Hua Chen}
  \author{Jian-Bo Deng} \email[Jian-Bo Deng: ]{dengjb@lzu.edu.cn}

  \affiliation{Institute of Theoretical Physics, LanZhou University,
    Lanzhou 730000, P. R. China}

  \date{\today}

  \begin{abstract}
   Many studies have been carried out since T.Padmanabhan proposed  that the cosmic acceleration can be understood from the perspective that spacetime dynamics is an emergent phenomenon. Motivated by such a new paradigm, we firstly study the de Sitter universe from emergence of space. After that we investigate the universes in general cases and then narrow down our discussions into one of them with a detailed discussion of the possibility in describing our real universe classically. Furthermore, a constraint on $Ht$ and a estimated value of $\tilde\Omega _{\Lambda}$ (caused by $\rho _{vac}$) can be derived from our model,  %The experimental data show the validity of our model and we can have some important conclusions by comparing with experimentas
the comparison with experiments is also presented. The results show the validity of our model.

  \end{abstract}


  \maketitle

  \section{Introduction}

  The discovery of black hole thermodynamics~\cite{1,2} has helped us to know the nature of gravity. With the deep study of the connection between gravitation and thermodynamics, physicists generally believe that the space is emergent which means gravity may not be a fundamental interaction but an emergent phenomenon now.

  It was first shown by Jacobson~\cite{3} that the Einstein field equations can be derived from the Clausius relation on a local Rindler causal horizon. Verlinde~\cite{4}, by suggesting that gravity should be explained as an entropic force caused by changes of entropy associated with the information on the holographic screen, put forward a great step towards understanding the nature of gravity. With the holographic principle and the equipartition law of energy, Verlinde derived the Newton's law and the Einstein fields equations in a relativistic regime. Earlier, Padmanabhan~\cite{5} observed that the equipartition law of energy for horizon degrees of freedom (DOF), combined with the thermodynamics relation $ S=\frac {E}{2T} $, leads to Newton's law of gravity.

  In most cases, only the gravitational field is treated as an emergent phenomenon, with the pre-exiting background geometric manifold assumed. A more complete way is to treat spacetime itself as an emergent structure as well, and it was finally proposed by Padmanabhan~\cite{6,7}. He argued that the spatial expansion of our universe is due to the difference between the surface DOF and the bulk DOF in the region of emerged space. Then, he proposed a simple equation $dV/dt=L^2_P\Delta N$~\cite{6,7}, where $V$ is the Hubble volume and $t$ is the cosmic time. $\Delta N=N_{sur}-N_{bulk}$ with $N_{sur}$ being the number of DOF on the boundary and $N_{bulk}$ being the number in the bulk.
Cai~\cite{cai2012emergence} generalized the derivation process to the higher (n+1)-dimensional spacetime. He also obtained the Friedmann equations of a flat FRW universe in Gauss-Bonnet and more general Lovelock cosmology by properly modifying the effective volume and the number of DOF on the holographic surface from the entropy formulae of static spherically symmetric black holes~\cite{cai2012emergence}. In Ref~\cite{yang2012emergence}, on the other hand, the authors generalized the holographic equipartition and derived the Friedmann equations by assuming that $(dV/dt)$ is proportional to a general function $f(\Delta N,N_{sur})$. Note that the authors of~\cite{cai2012emergence,yang2012emergence} only derived the Friedmann equations of the spatially flat FRW universe. In Ref~\cite{sheykhi2013friedmann}, Sheykhi derived the Friedmann equations of the FRW universe with any spatial curvature. The authors of~\cite{ai1} proposed a general equation which can be reduced to the different modified ones in different cases. For more investigations about the novel idea see Refs.~\cite{sheykhi2013friedmann,tu2013emergence,ai1,ai2,Sepehri:2015jaa,Sepehri:2014jla}.\par

  In this paper, we use the equation proposed by Padmanabhan to find some characters of the de Sitter universe at first, and then we generalize the important character $\rho+3p=constant$ to general case. By solving the equation $dV/dt=L^2_P\Delta N$, we can get the solution of $H(t)$, hence $a(t)$. Therefore we build a cosmological model to study the evolution of $a(t)$ in detail. It is easy to find that our universe would be de Sitter universe far into the future and a constraint on $H$ and $t$ is obtained in our model. Finally, we give a estimated value of $\tilde\Omega _{\Lambda}$(caused by $\rho _{vac}$) and compare them with the experiments data. This paper is organized as follows: In Section \uppercase\expandafter{\romannumeral 2}, a brief review of Padmanabhan's work is presented firstly. In section \uppercase\expandafter{\romannumeral 3}, we then discuss a de Sitter universe from emergence of space. In  Section \uppercase\expandafter{\romannumeral 4} we present our cosmological model in details and the comparison of our model with experiments. Section \uppercase\expandafter{\romannumeral 5} is for conclusions and discussions.

  \section{emergence of space}
Padmanabhan~\cite{6} noticed that in a pure de Sitter universe with Hubble constant $H$~, the holographic principle can be expressed in terms of
\begin{equation}
\label{1} N_{sur}=N_{bulk},
\end{equation}
where $N_{sur}$ denotes the number of DOF on the spherical surface of Hubble radius $ H^{-1} $, namely $ N_{sur}=4\pi H^{-2}/L_p^2 $, with $ L_p $ being the Planck length, while the bulk DOF $ N_{bulk}=\left|E\right|/(1/2)T $. Here $ \left|E\right|=\left|\rho+3p\right|V $, is the Komar energy with the Hubble volume $V=4\pi /(3H^3)$ and the horizon temperature $ T=H/2\pi $. For the pure de Sitter universe, substituting $ \rho=-p $ into Eq.~\eqref{1}, the standard result $ H^2=8\pi L_p^2\rho/3 $ is obtained.

From Eq.~\eqref{1}, one can get $|E|=(1/2)N_{sur}T$, which is the standard equipartition law. Padmanabhan called it \emph{holographic equipartition}, because it relates the effective DOF residing in the bulk to the DOF on the boundary surface. It is known that our real universe is just asymptotically de Sitter. Padmanabhan further suggested that the emergence of space occurs and relates to the difference $\Delta N=N_{sur}-N_{bulk}$. A simple equation was proposed~\cite{6}
\begin{equation}
\label{2} \frac{dV}{dt}=L_p^2\Delta N.
\end{equation}
Putting the above definition of each term, one obtains
\begin{equation}
\label{3} \frac{\ddot{a}}{a}=-\frac{4\pi L_p^2}{3}(\rho +3p).
\end{equation}
This is the standard dynamical equation for the FRW universe in general relativity. Using continuity equation $\overset{.}{\rho}+3H(\rho+p)=0$, one gets the standard Friedmann equation
\begin{equation}
\label{4} H^2+\frac{k}{a^2}=\frac{8\pi L_p^2\rho}{3},
\end{equation}
where $k$ is an integration constant, which can be interpreted as the spatial curvature of the FRW universe. Here, Padmanabhan takes $(\rho+3p)<0$, which makes sense only in the accelerating phase. It means that in order to have the asymptotic holographic equipartition, the existence of dark energy is necessary.

  \section{de Sitter universe from emergence of space}
As Padmanabhan said, this new idea provides a new paradigm for cosmology. Hence we would like to push an investigation on cosmology from the emergence of space. First, we will begin with the de Sitter universe.

In this case, the Eq.~\eqref {2} has the form
 \begin{equation}
\label{5} \frac{dV}{dt}=L_p^2(N_{sur}-N_{bulk})=0.
\end{equation}
One can easily obtain
\begin{equation}
\frac{dV}{dt}=\frac{d(\frac{4\pi}{3H^3})}{dt}=0,
\end{equation}
which means $H$ is a constant, and then $T$, $V$, $N_{sur}$, $N_{bulk}$ and  $|E|$ are all constants respectively.
Using $|E|=|\rho+3p|V$,  one can easily have
\begin{equation}
\label{assume}
|\rho+3p|= \frac{3H^2}{4\pi L_p^2}=constant
\end{equation}
in de Sitter universe. In other words, once $H$ is a constant, what we can have is $|\rho+3p|= constant$ according to emergence of space and $\rho = -p$ is just one special kind of de Sitter universe. So it is natural to generalize $p=-\rho$ to the general equation of state (EOS) $p=\omega\rho$~(~$\omega$ can be time dependent).

 %Noticing that the energy density for matter $\rho _M$: $\rho _M \propto a^{-3}$ and the energy density for radiation $\rho _R$: $\rho _R \propto a^{-4}$, we assume that the total energy density (may include vacuum energy density $\rho _{vac} $ and other unknown source )  $\rho=\rho (a)$ which means $\rho$ is only dependent on $a$. Using continuity equation $d(\rho a^3)/da =-3pa^3$ and our assumption $\rho=\rho (a)$, one can have $p=p(a)$. In this case, Eq. \eqref {assume} would be the term
%\begin{equation}
%\label{assume1}
%\rho(a)+3p(a)=-\frac{3H^2}{4\pi L_p^2}=-B_1,
%\end{equation}
%here we take the accelerating phase $\rho(a)+3p(a)<0$.
If we take the accelerating phase $\rho+3p<0$. then  Eq.~\eqref {assume} would be the term
\begin{equation}
\label{assume1}
\rho+3p=-\frac{3H^2}{4\pi L_p^2}= constant =-B_1,
\end{equation}

In de Sitter universe, combining Eq.~\eqref {assume1} and continuity equation $\overset{.}{\rho}+3H(\rho+p)=0$, one can have solutions of $\rho$ and $p$
\begin{equation}
\label{solution of rho}
\rho= \begin{cases}
\frac{1}{2}a^{-2}+\frac{B_1}{2},  &-1<\omega <-1/3 ~(\overset{.}{\rho} <0)\\
\frac{B_1}{2},&\omega =-1 ~(\overset{.}{\rho}=0)\\
\frac{B_1}{2}-\frac{1}{2}a^{-2}, &\omega <-1 ~(\overset{.}{\rho}>0)
\end{cases},
\end{equation}
\begin{equation}
\label{solution of p}
p= \begin{cases}
-\frac{1}{6}a^{-2}-\frac{B_1}{2},  &-1<\omega <-1/3 ~(\overset{.}{\rho} <0)\\
-\frac{B_1}{2},&\omega =-1 ~(\overset{.}{\rho}=0)\\
-\frac{B_1}{2}+\frac{1}{6}a^{-2}, &\omega <-1 ~(\overset{.}{\rho}>0)
\end{cases},
\end{equation}
where $a=Ae^{Ht}$ has absorbed the integral constant. From Eq.~\eqref{solution of rho} and Eq.~\eqref{solution of p}, we can see that there are strong constraints on $\rho$ and $\omega$ in de Sitter universe. Such results motivate us to move on to a cosmological model in a more general case.

\section{A cosmological model from emergence of space}
As well known, our real universe will evolve asymptotically to the de Sitter which satisfies the condition of $\rho+3p=constant$ as demonstrated in Section III. This naturally moves us to wonder whether the de-Sitter universe is the only case under this condition or other possibilities might exist and, if they do, what will they behave like. %We think this is quite a good motivation to rise our study.
In other words, does the equation
\begin{equation}
\label{assume2}
\rho + 3p =\text{constant}=-B_2
\end{equation}
only apply to de Sitter universe? Apparently not, Eq.~\eqref {assume2} can be obtained in the universe whose
\begin{equation}
\label{solution of rho1}
\rho= \begin{cases}
\frac{1}{2}a^{-2}+\frac{B_2}{2},  &-1<\omega <-1/3 ~(\overset{.}{\rho} <0)\\
\frac{B_2}{2},&\omega =-1 ~(\overset{.}{\rho}=0)\\
\frac{B_2}{2}-\frac{1}{2}a^{-2}, &\omega <-1 ~(\overset{.}{\rho}>0)
\end{cases},
\end{equation}
\begin{equation}
\label{solution of p1}
p= \begin{cases}
-\frac{1}{6}a^{-2}-\frac{B_2}{2},  &-1<\omega <-1/3 ~(\overset{.}{\rho} <0)\\
-\frac{B_2}{2},&\omega =-1 ~(\overset{.}{\rho}=0)\\
-\frac{B_2}{2}+\frac{1}{6}a^{-2}, &\omega <-1 ~(\overset{.}{\rho}>0)
\end{cases}.
\end{equation}
%where~$-B_2$~($B_2>0$)~is the constant in Eq.~\eqref {assume2}.

%What would this kind of universe be like in general?

%Now we will consider the universe where Eq. \eqref {assume2} is satisfied generally, which can be treated as a generalization of de Sitter universe and possibly represent our universe.

 To investigate the general case of this kind of universe, we go back to solve the Eq.~\eqref {2} with Eq.~\eqref {assume2}
\begin{equation}
\label{9}
\frac{dV}{dt}=\frac{d(\frac{4\pi }{3H^3})}{dt}=-\frac{4\pi}{H^4}\dot{H},
\end{equation}
\begin{equation}
\label{10}
L^2_{p}(N_{sur}-N_{bulk})=L^2_{p}(\frac{4\pi }{L^2_{p}H^2}-\frac{16\pi ^2B_2}{3H^4}).
\end{equation}
Combining and arranging the above equations, one can have
\begin{equation}
\label{11}
\alpha ^2-H^2=\frac{dH}{dt},
\end{equation}
where $\alpha=\sqrt{\frac{4\pi B_2 L^2_p}{3}}$. The solutions of Eq.~\eqref {11} are following($t>0$):

1):  $0<H<\alpha,~ dH/dt>0$
\begin{equation}
\label{12}
H=\alpha -\frac{2\alpha}{C_1e^{2\alpha t}+1},
\end{equation}
where $C_1$ is an integral constant satisfying $C_1\geqslant 1$ for the request of $H>0$.

2):  $H=\alpha, ~dH/dt=0 $
\begin{equation}
\label{13}
H=\alpha
\end{equation}

3):  $H>\alpha,~dH/dt<0$
\begin{equation}
\label{14}
H=\frac{2\alpha}{D_1e^{2\alpha t}-1}+\alpha,
\end{equation}
where $D_1$ is an integral constant satisfying $D_1\geqslant 1$ for the request of $H>\alpha$.

4): $-\alpha<H<0,~dH/dt>0$
\begin{equation}
H=\alpha -\frac{2\alpha}{M_1e^{2\alpha t}+1},
\end{equation}
where $M_1$ is an integral constant satisfying~$0<M_1<e^{-2\alpha t}$~ for the request of $-\alpha<H<0$~.

5): $H=-\alpha, ~dH/dt=0 $
\begin{equation}
H=-\alpha
\end{equation}

6):  $H<-\alpha,~dH/dt<0$
\begin{equation}
H=\alpha -\frac{2\alpha}{1-N_1e^{2\alpha t}},
\end{equation}
where $N_1$ is an integral constant satisfying~$0<N_1<e^{-2\alpha t}$~ for the request of $H<-\alpha$.

It is obvious that the first three solutions represent expansion and the last three represent contraction. Considering the fact our universe is expanding, one should be interested in the first three solutions, so we have a follow-up study.

According to $H=\dot{a} /a$, we have $a(t)$:

1): $0<H<\alpha$
\begin{equation}
\label{15}
a=\frac{C_1e^{2\alpha t}+1}{C_2e^{\alpha t}},
\end{equation}
where $C_2$ is an integral constant satisfying $C_2>0$.

2): $H=\alpha $
\begin{equation}
\label{16}
a=F_1e^{Ht},
\end{equation}
where $F_1$ is an integral constant satisfying $F_1>0$.

3): $H>\alpha$
\begin{equation}
\label{17}
a=\frac{D_1e^{2\alpha t}-1}{D_2e^{\alpha t}},
\end{equation}
where $D_2$ is an integral constant satisfying $D_2>0$.

 Next, let us push forward with a more detailed analysis on these three cases respectively.

1):  $0<H<\alpha,~ dH/dt>0$
\begin{equation}
H=\alpha -\frac{2\alpha}{C_1e^{2\alpha t}+1} ~, ~~ a=\frac{C_1e^{2\alpha t}+1}{C_2e^{\alpha t}},
\end{equation}
where $C_1\geqslant 1$ and $C_2>0$.
In the limit of $t\rightarrow 0$, one can have $a_0=(C_1+1)/C_2 > 0 $. This means a universe has an initial nonzero scale factor $a_0$. %%With the normal initial condition $a(0)=0$ imposed, it is not proper to be the evolution of a real universe.

2):  $H=\alpha, ~dH/dt=0 $
\begin{equation}
H=\alpha,~a=F_1e^{Ht},
\end{equation}
where $F_1>0$. This is exactly the de Sitter uninerse which we have discussed in Section \uppercase\expandafter{\romannumeral 3}.

3):  $H>\alpha,~dH/dt<0$
\begin{equation}
\label{28}
 H=\frac{2\alpha}{D_1e^{2\alpha t}-1}+\alpha~, ~a=\frac{D_1e^{2\alpha t}-1}{D_2e^{\alpha t}},
 \end{equation}
where $D_1\geqslant 1$ and $D_2>0$. In the limit of $t\rightarrow 0$, $a_0=(D_1-1)/D_2$. If one set $D_1=1$, then $a_0=0$.

So far, we have got all the universe where Eq.~\eqref {assume2} is satisfied generally, and de Sitter universe is just one case of them as we predicted. It is natural to wonder whether our real universe could be one of them.

For our universe, $a_0=0$. So, we would like to set $D_1=1$ in Eq.~\eqref {28} and we can have
\begin{equation}
\label{18}
H=\frac{2\alpha}{e^{2\alpha t}-1}+\alpha,
\end{equation}
\begin{equation}
\label{19}
a=\frac{e^{2\alpha t}-1}{D_2e^{\alpha t}}=A(e^{\alpha t}-\frac{1}{e^{\alpha t}}),
\end{equation}
where $A=1/D_2 >0$.
%Then it will have the possibility to describe our universe. %%Therefore we will take it as a cosmological model from emergence of space and give a detailed investigation.
%%We are not sure that Eq.~\eqref {19} does represent our real universe. However, there is no special reason it can not represent our universe, and we will take it as a cosmological model from emergence of space and give a detailed investigation.
It is easy to find that Eq.~\eqref {19} describes a universe which is asymptotically de Sitter and our universe is in this case as we generally believe.

There are two problems if Eq.~\eqref {19} is used to describe our universe. First,  Eq.~\eqref {19} is obtained from the universe which satisfies Eq.~\eqref {assume2}. It is natural to doubt that our universe satisfies Eq.~\eqref {assume2}. However, it may be correct considering that our late universe would be de Sitter which satisfies Eq.~\eqref {assume2}. Second,
it can not explain the inflation of early universe. However, just as Padmanabhan~\cite{6} said, Eq.~\eqref {2} needs modifications at early universe. Except these two problems, there is no obvious reason to sweep out the possibility of our real universe depictured by Eq.~\eqref{19}. So, we will take it as a cosmological model from emergence of space and give a detailed investigation.

%Since $t>0$ and $\alpha>0$, there should be a constraint on $H$ (hence $a$). To derive this constraint, we will reverse the process to solve $\alpha$ by $H$ and $t$. We are going to get the constraint under which the $\alpha>0$ exists.
Since $t>0$ , $\alpha>0$~and $H>\alpha$, there should be a constraint in Eq.~\eqref {18}. To find out the constraint, one can multiply $t$ on both sides of Eq.~\eqref {18} and substitute $x=\alpha t$, then one can have
\begin{equation}
\label{Ht}
Ht=\frac{2x}{e^{2x}-1}+x  \text{\ \  $(x>0)$.}
\end{equation}
Let \[y=\frac{2x}{e^{2x}-1}+x \text{\ \  $(x>0)$}.\]
After calculating, one can find
\[y'>0~,~\underset {x\rightarrow 0}{\lim}~y=1 ~.\]
So the existence of  a solution of $\alpha>0$ requires a constraint on $H$ and $t$:
\begin{equation}
\label{constrain}
tH(t)>1 \text{  ~or~~} t>1/H(t).
\end{equation}

Eq.~\eqref {constrain} is applicable for any $t>0$ (except for $t\rightarrow 0$ which represents the inflation of early universe).
Then one can have
\begin{equation}
\label{constrain1}
t_0H_0>1,
\end{equation}
where $t_0$ is the present age of our real universe and $H_0$ is the current value of $H$.%ubble parameter%

%One may find that $t_0H_0$ (which was less than one as we generally believed before the observations of Type Ia Supernovae) can be greater than one when we find the Supernovae for an accelerating Universe.

Since the constraint is only derived in our model and has never appeared in other theories before as far as we know, we would like to compare it with the experiments~\cite{8,9,10,11,12} from the Wilkinson Microwave Anisotropy Probe (WMAP) and the Planck Mission. The analysis is shown in Table I.
\clearpage
\begin{table}[h]
\newcommand{\tabincell}[2]{\begin{tabular}{@{}#1@{}}#2\end{tabular}}
\caption{An analysis of the data of $H_0$ and $t_0$ from the WMAP and the Planck Mission. For obtaining $H_0t_0$, we have changed km/(Mpc$\cdot$s) into $s^{-1}$ and Ga into $s$. %Using the Eq.~\eqref{18}, we have got the maximum~$\alpha$ for each case.
} % title of Table
\centering % used for centering table
\begin{tabular}{c c c c c c c} % centered columns (4 columns)
\hline\hline %inserts double horizontal lines
Obsever & \tabincell{c}{Data published} ~~& ~~\tabincell{c}{$H_0$\\km/(Mpc$\cdot$s)} ~~&~~ \tabincell{c}{$t_0$(Ga)} ~~&~~~~~~~$H_0t_0~~~~~~~~~~$ & \tabincell{c}{$\alpha$\\km/(Mpc$\cdot$s)}\\ [0.5ex] % inserts table
%heading
\hline % inserts single horizontal line
WMAP &2003  & $71_{-3}^{+3}$ & $13.7_{-0.2}^{+0.2}$ & 0.9393$\sim$1.0667 & 31.66 &\\ % inserting body of the table
WMAP & 2006 & $73.2_{-3.2}^{+3.1}$ & $13.73_{-0.15}^{+0.16}$& 0.9727$\sim$1.8044 & 35.71& \\
WMAP & 2008 & $70.5_{-1.3}^{+1.3}$ & $13.72_{-0.12}^{+0.12}$& 0.9629$\sim$1.0168  & 16.86 &\\
WMAP &2010 & $70.4_{-1.3}^{+1.3}$ & $13.75_{-0.11}^{+0.11}$ & 0.9644$\sim$1.0168 &15.90&\\
Planck& 2013 & $67.80_{-0.77}^{+0.77}$ & $13.798_{-0.037}^{+0.037}$&0.9438$\sim$0.9707 & no value
\\ [1ex] % [1ex] adds vertical space
\hline %inserts single line
\end{tabular}
\label{table1} % is used to refer this table in the text
\end{table}

Here, we would like to have an explanation of the value of $\alpha$ in Table I. According to Eq.~\eqref{18}, there exists a value of $\alpha$ in our model once $H_0t_0>1$. And the value of $\alpha$ we have calculated in Table I is the maximum one in each case.

All the experimental data in Table I show that $H_0t_0\approx 1$ and  some experiments have $H_0t_0>1$. Noticing that the experimental date which $H_0t_0<1$ depart from one by $10^{-2}$, our model is fairly valid to describe the real universe.

In fact, our theory may provide a new judgment of $H_0t_0>$1 for the experimental data, which is indeed supported by most of the experimental data from WMAP. And the departure from $H_0t_0>$1 can be understood well as a uncertainty of current measurement accuracy.

%However, by comparing the data from the WMAP and the Planck, our model tends to support the WMAP rather than the Planck. \\

%The above results have shown the availability of our model in describing the real universe.
What is more, we might actually calculate the vacuum energy if our model is used to describe the real universe. To see this, let us go back to Eq.~\eqref{solution of rho1}
 \[
\rho= \begin{cases}
\frac{1}{2}a^{-2}+\frac{B_2}{2},  &-1<\omega <-1/3 ~(\overset{.}{\rho} <0)\\
\frac{B_2}{2},&\omega =-1 ~(\overset{.}{\rho}=0)\\
\frac{B_2}{2}-\frac{1}{2}a^{-2}, &\omega <-1 ~(\overset{.}{\rho}>0)
\end{cases}
\]
where $B_2=3\alpha ^2/(4\pi L_p^2)$ and $a$ is given in the form as in Eq.~\eqref{15}$\sim$\eqref{17},\eqref{19} respectively. %Though it can be read from Eq.\eqref{solution of rho1} that the second line in the RHS should correspond to the vacuum energy, we still would like to give a more confident argument.
 Even though it is impossible to confirm which one ($\dot \rho <0 $ or $\dot \rho =0$ or $\dot \rho >0$) belongs to the universe described by Eq.\eqref{19}, one may find that the $\rho$ of our real universe would be $\dot \rho <0 $ by noticing that the energy density for matter $\rho _M$: $\rho _M \propto a^{-3}$ and the energy density for radiation $\rho _R$: $\rho _R \propto a^{-4}$. Hence we should choose the first line in the RHS of Eq.\eqref{solution of rho1} to be the possible energy content of our universe.

With the energy content given in the form
\begin{equation}
\label{the rho of our universe}
\rho =\frac{1}{2}a^{-2}+\frac{B_2}{2}~,
\end{equation}
it is natural to inquire the possible meaning of $B_2/2$~. Noticing that $\underset {a\rightarrow +\infty }{\lim}~\rho =B_2/2=-p$~is the vacuum energy density($\rho _{vac}$) of the pure de Sitter which our real universe would be in the far future, we argue that $B_2/2$ is the $\rho _{vac}$ of our real universe and
\begin{equation}
\label{rho0}
\rho _{vac}=\frac{B_2}{2}=\frac{3\alpha ^2}{8\pi L_p^2}.
\end{equation}

By dimensional analysis, we have

\[[\frac{3\alpha ^2}{8\pi L_p^2}]=\frac{[T]^{-2}}{[L]^{2}}~,~[\rho]=\frac{[M]}{[L]^3}=\frac{[T]^{-2}}{[L]^{2}}\cdot \frac{[T]^2[M]}{[L]}.\]
We now put back the fundamental constants and get
\[\rho _{vac}=\frac{3\alpha ^2}{8\pi L_p^2}\cdot \frac{t_p^2m_p}{L_p}~~~.\]

Since $\alpha$ can be calculated by experimental data $H_0$ and $t_0$ in our theory, we are actually building a new relation of $H_0$, $t_0$ and vacuum energy which has never appeared in the literature.  Using the $\alpha$ in the Table I, one can finally obtain the range of $\rho _{vac}$~:
\begin{equation}
4.733\times 10^{-28}\sim 2.397\times 10^{-27}~kg/m^3.
\end{equation}

%~\cite{Tegmark:2003ud}-26
Comparing with the density of dark energy ( about $6.91\times 10^{-27} kg/m^3$)~\cite{8,9,10,11}, there is a difference between the two though it almost has the same order of our estimated value.

%In fact, the difference can be showed in Table I directly.  %So we can not jump to a conclusion that the form for dark energy content is $\rho _{vac}$ .

%We know ~\cite{Carroll:2000fy} that two proposed forms for dark energy $\Omega _{\alpha}$  are the cosmological constant $\alpha$ or the vacuum energy density, a constant energy density filling space homogeneously, and scalar fields such as quintessence or moduli, dynamic quantities whose energy density can vary in time and space.
In the standard $\Lambda$CDM cosmological model, it is believed that the dark energy is caused by the cosmological constant. Hence it is convenient to compare our theoretical results with the experiment data of $\Omega_{\Lambda}$.

The definition of $\Omega_{\Lambda}$ is
\begin{equation}
\Omega _\Lambda = \Lambda/(3H_0^2)~,~~
\end{equation}
where $\Lambda =8\pi \rho _{vac}$.
Using the $\rho _{vac}$ in our model, one can have
\begin{equation}
\label{31}
\tilde{\Omega} _{\Lambda} = \alpha ^2/H_0^2 ~~.
\end{equation}
According to the values of $\alpha$ and $H_0$ in Table I, one can finally get the range of $\tilde\Omega _{\Lambda}$ :
\begin{equation}
0.049 \sim 0.260~~~~.
\end{equation}
And the experimental data of $\Omega _{\Lambda}$ from the WMAP and the Planck Mission~\cite{8,9,10,11,12} have the range of :
\[0.683\sim 0.772~~~.\]

It can be seen that our model predicts the approximate but not exact value of cosmological constant. Still, the cosmological constant derived in our model has the same order of the experimental data. The difference may indicate that there are possibly other sources of dark energy such as the quintessence if our model represents the real universe.
%A notable difference between the two is easy to find. So, our model tends to support that the form for most of the dark energy would be scalar fields at present.

%In fact, from Eq .\eqref{18}
%From Eq .\eqref{the rho of our universe}, one can find that there should be other unknown source whose $\hat \rho \propto a^{-2} $  at least by noticing that the energy density for matter $\rho _M$: $\rho _M \propto a^{-3}$ and the energy density for radiation $\rho _R$: $\rho _R \propto a^{-4}$.

\section{Conclusions And Discussions}
 To summarize, in this paper,  we investigated the novel idea proposed by Padmanabhan~\cite{6} that the emergence of space and expansion of the universe are due to the difference between the number of DOF on the holographic surface and the one in the emerged bulk. It is shown that the Friedmann equation of a flat FRW universe can be derived  with the help of continuity equation. Since the emergence of space may provide a completely different paradigm to study cosmology~\cite{6}, we studied the de Sitter universe from emergence of space, and found that there is a constraint on $\rho$ and $p$ ( Eq.~\eqref{assume} and Eq.~\eqref{assume1}) which can derive solutions of $\rho$ and $p$~(Eq.~\eqref{solution of rho} Eq.\eqref{solution of p}). By considering an arbitrary universe whose $\rho$ and $p$ have the form of Eq.\eqref {solution of rho1} and Eq.\eqref {solution of p1}, we generalized Eq.\eqref{assume1} beyond the de Sitter universe and solved Eq.\eqref {2}.

 Among the solutions we obtained, we found a model which has the possibility to describe our real universe. After detailed analysis of our model, we got three important conclusions:
 \begin{itemize}
\item[(1)] The universe would be de Sitter in its later period. ($t>> 1/\alpha $).
\item[(2)] There is a constraint on $H$ and $t$~:~$H(t)\cdot t>1$, and it is applicable for any $t>0$ (except for $t\rightarrow 0$ which represents the inflation of early universe).
\item[(3)] The value of vacuum energy and $\tilde\Omega _{\Lambda}$ can be derived in our model.
\end{itemize}

We made a comparison of our model with experiments. For conclusion (2), the experimental data show that $H_0t_0$ ranges from $0.9438\sim 1.8044$ and our model tends to support the WMAP rather than the Planck. For conclusion (3), our model predicts a positive tiny cosmological constant, which is approximate to the experimental data. The difference may indicate that there are probably other sources contributing to the dark energy if our model represents the real universe.

%It is natural to doubt that our universe has Eq.~\eqref {assume2}. However, it may be correct considering that our late universe would be de Sitter which has Eq.~\eqref {assume2}.
%Though we did not assume that our universe has Eq.~\eqref {assume2} at first,
 %Finally, we would like to provide an account of applying Eq.~\eqref {assume2} to our universe. we just found all the universe where Eq.~\eqref {assume2} is satisfied generally. Among the universe we have obtained, we found a possible model to describe our universe. And the results of the comparison with experiments show the validity of our model in describing the real universe.

\section*{ACKNOWLEDGMENTS}

  We would like to thank the National Natural Science Foundation of
  China~(Grant No.11171329) for supporting us on this work.

  \bibliographystyle{apsrev4-1}
  \bibliography{reference}

\end{document}